\documentclass[]{article}

\usepackage{arxiv}

\usepackage[utf8]{inputenc} 
\usepackage[T1]{fontenc}    
\usepackage{hyperref}       
\usepackage{url}            
\usepackage{booktabs}       
\usepackage{amsfonts}       
\usepackage{nicefrac}       
\usepackage{microtype}      
\usepackage{lipsum}
\usepackage{graphicx}
\usepackage{natbib}
\usepackage{index, amsmath, listings, multirow, nomencl, ltablex, longtable,  rotating, scalerel,  setspace, subfig,tabularx, titlesec,url, verbatim, adjustbox, algpseudocode, float, listings, multirow, nomencl, ltablex, longtable,  rotating, scalerel, setspace, subfig, tabularx, bm, makecell, hyperref, natbib}

\title{Component over Composite: Mitigating Type I Error Inflation when Imputing ``Days Alive and at Home''}

\author{
Mia S. Tackney \\
  MRC-Biostatistics Unit, University of Cambridge, Cambridge, United Kingdom  \\
  \texttt{mst35@cam.ac.uk} \\
   \And
  Sarah Dawson  \\
 MRC-Biostatistics Unit, University of Cambridge, Cambridge, United Kingdom  \\
  Papworth Trials Unit Collaboration, Royal Papworth Hospital, Cambridge, UK\\
  \And
 Letao Yuan \\
 MRC-Biostatistics Unit, University of Cambridge, Cambridge, United Kingdom  \\  
  \And 
 Dominique-Laurent Couturier \\
 MRC-Biostatistics Unit, University of Cambridge, Cambridge, United Kingdom  \\
 Papworth Trials Unit Collaboration, Royal Papworth Hospital, Cambridge, UK\\
 \And 
Sofía S. Villar \\
 MRC-Biostatistics Unit, University of Cambridge, Cambridge, United Kingdom  \\
 Papworth Trials Unit Collaboration, Royal Papworth Hospital, Cambridge, UK\\
}

\begin{document}
\maketitle
\begin{abstract}
\textbf{Background} Days Alive and at Home (DAH) over a pre-defined follow-up period is a novel post-intervention composite outcome that combines data from at least three components: (i) initial length of hospital stay, (ii) length of total readmissions or other post-discharge care and (iii) mortality. Missing values bring unique challenges to the analysis of trials with the DAH outcome as the three components may have different rates of missingness caused by distinct missing data mechanisms. Current approaches define DAH as missing if any of the components are missing, and proceed with complete cases or Multiple Imputation (MI) of the composite. \textbf{Methods} Through a simulation study motivated by the NOTACS trial, we compare several methods of handling missing data, including complete case analysis, MI of the composite, and MI of the components when the primary analysis is a Mann-Whitney-Wilcoxon test. \textbf{Results} MI on the component level has good properties in terms of type I error control and power. We caution against the use of MI on the composite level with Predictive Mean Matching, which can lead to type I error inflation. \textbf{Conclusions} Given the complex distributional characteristics of DAH, naive approaches such as defining missingness on the composite level and directly imputing the composite with Predictive Mean Matching, can lead to type I error inflation. Imputing on the component level is recommended, suggested future work included imputation approaches that are compatible with more complex definitions of DAH, as well as recommendations for sensitivity analyses to the Missing at Random assumption.   
\end{abstract}


\section{Introduction}
A novel, patient-centric outcome which is increasingly being used in post-operative trials is \textit{Days Alive and at Home} within a follow-up period such as 90 (DAH90) or 30 (DAH30) days \citep{Bell2019}. As a composite outcome, DAH combines data from at least three components: (i) the length of the initial post-intervention hospital stay, (ii) total length of hospital re-admissions or other care in the follow-up period, and (iii) mortality \citep{Myles2017}. Missing data presents unique challenges for analyses of RCTs with composite outcomes such as DAH. Data from each component may be collected via a different process, leading to components which have different rates of missingness and different missing data mechanisms. For DAH, the length of initial hospital stay and mortality, typically collected at trial centres and relying on hospital data, are generally complete, whereas participants' locations after the initial hospital stay often rely on participant location diaries and are therefore more prone to being missing. To retain data from observed components, and to reflect potentially different missing data mechanisms in different components, previous work has shown that performing Multiple Imputation (MI) at the component level is preferred over MI at the composite level \citep{OKeeffe2016, Pham2021, Ibrahim2020, Gachau2021}. However, these works have focused on composite outcomes that are binary or approximately normal and analysed via a regression model. As DAH has complex distributional characteristics \citep{YuanPLACEHOLDER} and is often analysed via a non-parametric test, the way missing data should be handled warrants further investigation. 

The Nasal Oxygen Therapy After Cardiac Surgery (NOTACS) trial is a recent phase III trial with DAH90 as the primary outcome \citep{KleinPLACEHOLDER}. It evaluated the efficacy and safety of high-flow nasal therapy (HFNT) compared to standard oxygen therapy (SOT) on participant outcomes after cardiac surgery. A total of 1280 participants were randomised on a 1:1 ratio to HFNT versus SOT (the initial sample size of 850 was updated after a sample-size re-estimation at interim). The primary analysis was a Mann-Whitney-Wilcoxon test of DAH90 by treatment group. Analysis plans have been published previously \citep{Dawson2022SAP, Dawson2024SAP_update, Shetty2025}. Of note, for the NOTACS trial, the definition of DAH90 was more complex than the one used by \cite{Myles2017}, as the total number of days without escalation of care was computed, rather than the total number of days at home irrespective of the level of care compared to baseline. This penalises, for example, days spent at home but with a higher level of care than at baseline. Thus, data on both location and level of support were required to compute DAH90 in the NOTACS trial. While missing DAH90 outcomes were minimial (only $5\%$), it serves as a motivating example which illustrates how missing data could arise and should be handled in other trials with the DAH outcome.

\begin{figure}[] 
    \centering
    \includegraphics[width=0.5\textwidth]{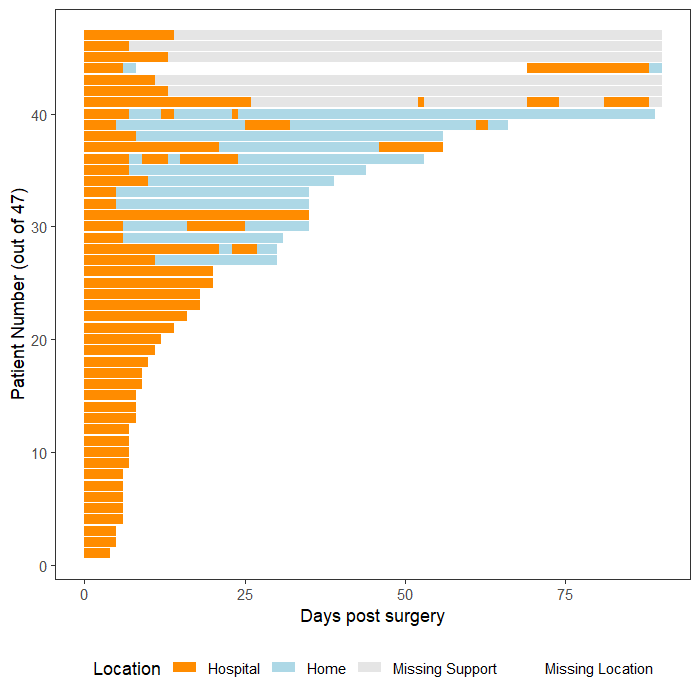}
    \caption{Colour-coded missingness status in location and support data from the 47 participants with missing data (y-axis) in the primary analysis for NOTACS for each day of the 90-day follow-up period (x-axis).}
    \label{location_support_missing}
\end{figure}

Figure \ref{location_support_missing} displays the missingness status for the location and support data used to compute DAH90 for participants who had missing values in the NOTACS trial (and did not withdraw consent). For all 47 participants shown, we observe that the data on the initial hospital stay were complete, as they were collected directly at the trial site. Data on mortality were also complete for the same reason; there were no deaths among the participants shown. Missingness occurred after hospital discharge, when patient-reported location diaries were used. Of the 47 participants shown, 27 dropped out after the initial hospital stay and did not engage with the location diaries; 13 dropped out after some time of using the location diaries and had missing location data. The remaining seven participants had missing data on support. Intermittent missingness occurs only in two instances. The high rate of completion overall, and the tendency of missing data to be on the tail end of the follow-up period, is likely a reflection of the effort of the trial team who sent frequent reminders. In other settings, it may be plausible that there would be more missing values overall, and more instances of intermittent missingness. 

An approach to handling missing data in DAH is to require a full record of the individual (requiring full location, support and mortality data during the follow-up period), and to perform complete case analysis \citep{Myles2018}. With this strategy, even one day where the participant's location is missing would deem the entire DAH observation as missing. In the NOTACS trial, the primary analysis proceeded with complete case analysis as there were $ \leq 5\%$ of participants with missing data. The statistical analysis plan stated that if the percentage of participants with missing data exceeded 5\%, this would require investigating which variables were associated with the missing data indicator, and, if there were any associations suggesting that these were Missing at Random (MAR) rather than Missing Completely at Random (MCAR), to use MI. In that case, an imputation model relating the composite outcome to the variables that predict missingess would be constructed separately for each treatment arm, and missing values imputed $M$ times via predictive mean matching. The Mann-Whitney-Wilcoxon test would be conducted on the $M$ completed datasets and the median of the $M$ p-values would be reported.

Trials rarely report the quantity of missing data for each component of DAH or how missing data is handled. A recent review of critical care trials using hospital-free days, an outcome closely related to DAH, revealed substantial inconsistencies in the reporting of individual components \citep{Shackleton2025} and noted that mortality was often not distinguished from other components. While DAH is rapidly gaining traction as a patient-centric outcome, critical gaps remain in understanding how missing data should be handled. The few studies which discuss missing data have defined missing on composite level, even though some components are fully observed. By handling missing data on the component level instead, the fully observed initial length of hospital stay can be used as a predictor of readmissions in the follow-up period. Further, as DAH has a complex distribution, typical imputation models may be misspecified and consequences of this have not been studied. This article serves to disentangle these complexities and provide a recommended approach to handling missing data with MI.

\section{Methods}
\subsection{Model for DAH90}
\label{sec_model}
\begin{figure}[] 
    \centering
    \includegraphics[width=0.5\textwidth]{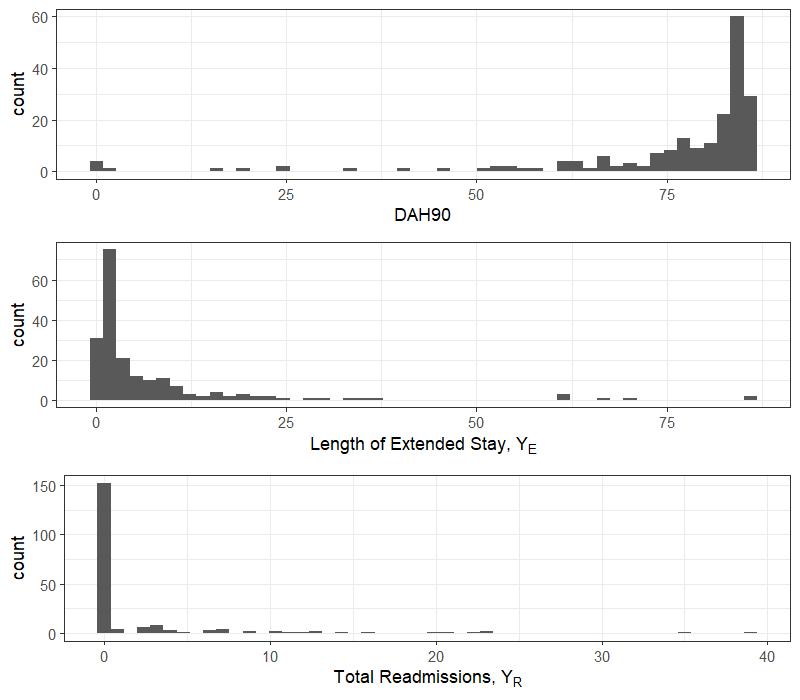}
    \caption{Top panel: Histogram of the DAH90 observations of a subset of 200 participants from the interim analysis of the NOTACS trial. Middle panel: Histogram of the length of extended post-intervention initial hospital stay for 198 participants who remained alive during the study; Bottom panel: Histogram of the total number of days spent in hospital following readmission(s) for the 198 participants who remained alive during the study.}
    \label{model_histograms}
\end{figure}

DAH90 has complex distributional properties and is characterised by zero-inflation, left-skewness, and bimodality. The top panel of Figure \ref{model_histograms} displays the histogram of  DAH90 observations of a subset of 200 participants from the interim analysis of the NOTACS trial. We observe that there is left-skewness, with most values above 75, as well as a small clump-at-zero both induced either by death (n=2) or initial hospital stays exceeding the 90-day period of interest (n=2). \\

The model we assume for DAH90 is the one developed by \cite{YuanPLACEHOLDER} using the NOTACS trial interim data, who decompose $y_i$, the DAH90 outcome for participant $i$, into its three components as follows:  

\begin{equation}
\label{DAH90_formula}
    y_i = [90 - (p+y_{E_i})-y_{R_i}] \mathbb{I}(d_i=0),
\end{equation}

where: 
\begin{enumerate}
    \item $p+y_{E_i}$ corresponds to the duration of the \textbf{post-intervention initial hospital stay} for participant $i$. This consists of a minimum length of stay, typically protocol-dependent, of $p$ days. In the NOTACS trial, $p=4$ days for all sites. For some participants, there was an extended stay of $y_{E_i}$ days, which may occur due to surgical complications or poor health. As illustrated in the middle panel of Figure \ref{model_histograms}, the distribution of $y_{E_i}$ is both right-skewed and right-censored due to long initial hospital stays. \cite{YuanPLACEHOLDER} found that the random variable $y_{E_i}$ was best modelled via a right-censored zero-inflated Poisson inverse Gaussian (ZICPIG) distribution for the NOTACS trial data. We denote by $v_i = p+y_{E_i}$ the total number of days in the post-intervention initial hospital stay for participant $i$, where $4 \leq v_i \leq 90$. 
    \item $y_{R_i}$ corresponds to the \textbf{number of days spent in readmission(s)} in the post hospital discharge period. As illustrated in the bottom panel of Figure \ref{model_histograms}, the random variable $y_{R_i}$ is zero-inflated and right-skewed, and \cite{YuanPLACEHOLDER} found this random variable was best modelled via a zero-adjusted beta binomial model (ZABB). We denote by $w_i=90-v_i$ the the number of days between hospital discharge and the end of study, which was used as binomial denominator for participant $i$.
    \item $d_i$ is a binary indicator of \textbf{mortality}, and takes value 1 if the $i$th participant died during the trial and 0 otherwise. This random variable was assumed to follow a Bernoulli distribution (with estimated probability of death equal to 0.01 based on the subset of the NOTACS interim data).
\end{enumerate}

 More details on this model are provided in Supplementary File 1.

\subsection{Methods for Handling Missing Data for an analysis of DAH90 with the Mann-Whitney-Wilcoxon Test}

\subsubsection{\textbf{Complete Case Analysis}}
If the composite outcome is likely to be MCAR, one common approach is to perform complete case analysis, where only participants who have all components fully observed are analysed. We note two points of caution; firstly, it can be difficult to ascertain the missing data mechanism of a composite, as even if each component is MCAR, it is possible for the resulting composite to be MNAR \citep{Pham2021}. Furthermore, for DAH90, in the case where a participant dies, their composite can be derived to take the value of zero even when other components are missing; however, when a participant remains alive in the study, full data on 90 days are required to compute the DAH90 outcome. \cite{Pham2021} demonstrated that, in such cases, deriving the composite can lead to bias and should be avoided. \\

\subsubsection{\textbf{Multiple Imputation (MI)}} 
Under a broader assumption of Missing at Random (MAR), MI can be used to handle missingness either at the composite or component level. For simplicity, we consider a setting in which only the DAH outcome has missing values, and treatment assignment and covariates are assumed to be fully observed. In this case, univariate MI can be performed separately by trial arm, as often recommended in clinical trial settings \citep{White2011, Sullivan2018}. \\

\subsubsection{\textbf{MI at the composite level}} 

Let $n$ denote the number of participants, $z$ denote the treatment allocation where $z=0$ and $z=1$ respectively correspond to the control and experimental arms, $Y$ denote the vector of length $n$ of DAH90 outcomes, and $X=\left[1,  X_1, ..., X_p \right]$ denote the matrix of covariates to be used in the imputation. Thus, $Y^z$ and $X^z$ respectively correspond to the outcome vector and covariate matrix of participants assigned to treatment $z$. The outcome $Y^z$ can further be partitioned into those that are observed and those that are missing: $Y^z=(Y^{z,obs}, Y^{z,mis})$. The number of participants assigned to treatment $z$ with observed outcomes is denoted $n^{z, obs}$ and covariate values corresponding to those with observed outcomes are denoted $X^{z, obs}$.\\

Then, univariate parametric MI proceeds as follows: 

\begin{enumerate}
    \item For each arm $z \in \left\{0, 1 \right\}$, fit an imputation model, assuming that the covariate-outcome relationship is linear:
    \begin{equation}
        Y^{z, obs} \mid  X^{z, obs} \sim \mbox{Normal}(X^{z, obs} \beta^z, \sigma^{z^2} I), 
    \end{equation}
    and compute estimates $\hat{\beta}^z$ and $\hat{\sigma}^{z^2}$.
    \item Draw from the posterior distribution of $\sigma^{z^2}$ and $\beta^z$ parameters under an uninformative prior: 
     \begin{align*}
        \tilde{\sigma}^{z{^2}} &\sim \frac{(n^{z, obs}-p) \hat{\sigma}^{z^2} }{\chi^2_{n^{z,obs}-p}},
    \end{align*}
    
    \begin{align*}
        \tilde{\beta}^{z} &\sim \mbox{Normal}\left(\hat{\beta}^z, \tilde{\sigma}^{z^2} \left(X^{{z, obs}^\top} X^{z, obs}\right)^{-1}\right)\\
    \end{align*}
    \item For missing outcomes, given their covariate values $x_i$, draw an imputation from the conditional predictive distribution: 
    \begin{equation}
        y_i^{imp} \sim \mbox{Normal}(x_i ^\top \tilde{\beta}^{z}, \tilde{\sigma}^{z{^2}}). 
    \end{equation}
    
    \item With the full dataset containing observed and imputed data from both arms, perform a Mann-Whitney-Wilcoxon test and obtain a p-value.
    \item Repeat steps 1-4 for a total of $M$ times. As Rubin's Rules \citep{Rubin1987} are not valid for a rank-based test statistic, we obtain $M$ p-values and report their median. We note that using the median p-value is an ad-hoc approach; while it is preferable to use a method that takes into account the between-imputation variance, this is beyond the scope of this investigation. 
\end{enumerate}

As the linearity assumption in Step 1 may not hold for the DAH outcomes or components, a popular approach in such cases is to use Predictive Mean Matching (PMM). Here, Step 3 is modified as follows: using the parameters drawn in Step 2, each missing outcome is assigned the DAH value of a donor who is randomly selected among the $k$ participants assigned to the same treatment arm that minimise the distance metric $\mid \hat{y}_i^z - x_i^\top \hat{\beta} \mid$. Here, where $\hat{y}_i^{z} = x_i ^\top \tilde{\beta}^z$ denotes the predicted DAH value of participant $i$. The default value of $k$, the number of donors, in statistical software such as \texttt{mice} is 5.

\subsubsection{\textbf{MI at the Component level}}
Instead of imputing the composite outcome directly, the missing values of the components of DAH are imputed. In Step 1, the component $y_{R_i}$ is imputed conditional on covariates as well as the fully observed component $y_{E_i}$. After generating an imputation $Y_{R_i}^{imp}$, in Step 4, the DAH90 value is reconstructed via Equation \eqref{DAH90_formula}, which is referred to as ``passive imputation" \citep{Seaman2012} or ``impute, then transform" \citep{vonHippel2009}. This process is repeated $M$ times. The reported p-value is the median of the $M$  p-values for the Mann-Whitney-Wilcoxon tests, each performed on one of the $M$ imputed datasets.

\subsection{Simulation Study}
\label{simulation}
We evaluated the statistical properties of several methods for handling missing outcomes when DAH90 has missing values in a setting motivated by the NOTACS trial. Assuming a total sample size of 1280 per simulation run, we generated key covariates (Age, Sex, Body Mass Index (BMI) and trial site) by bootstrapping these from a subset of 200 participants who were in the interim analysis. Table \ref{table_covar} displays the distribution of two key predictors by treatment arm in the interim data. Treatments were assigned with a 1:1 ratio. DAH90 values were then generated via the model by \cite{YuanPLACEHOLDER}. The probability of death is assumed to be 0.01. We considered two scenarios: under the Alternative scenario, which was powered to detect a two-day difference in medians between treatment groups with 90\% probability, and under the Null scenario, where we set treatment-related parameters to zero (see Supplementary File 1 for full details). \\

\begin{table}[H]
\centering
\begin{tabular}{llll}
  \hline
  & Standard & HFNO & Overall \\ 
  \hline
 & (N=102) & (N=96) & (N=198) \\ 
  \textbf{Sex} &  &  &  \\ 
    Male & 73 (71.6\%) & 60 (62.5\%) & 133 (67.2\%) \\ 
    Female & 29 (28.4\%) & 36 (37.5\%) & 65 (32.8\%) \\ \hline
  \textbf{Age} &  &  &  \\ 
    51- yo & 7 (6.9\%) & 11 (11.5\%) & 18 (9.1\%) \\ 
    51+ yo & 95 (93.1\%) & 85 (88.5\%) & 180 (90.9\%) \\ 
   \hline
\end{tabular}
\caption{Breakdown of covariates Sex and Age by treatment group in the interim analysis.}
\label{table_covar}
\end{table}

\subsection*{Missing Data Mechanism}
Let $S_i$ denote an ordinal variable indicating the missingness status of the post hospital discharge period, with levels \textit{Complete, Partial}, or \textit{Missing}. While the NOTACS trial had minimal missing data, we consider five possible missing data mechanisms that may occur in general: (1) MCAR, (2) MAR depending on Age, (3) MAR depending on Sex, (4) MAR depending on Age and Sex and (5) MNAR depending on Age. \\

We modelled $S_i$ using a multinomial logistic regression. Under MCAR, the probability of the post hospital discharge period being complete, partially observed or missing were fixed at 0.95, 0.03 and 0.02, respectively, regardless of covariate values. For MAR and MNAR mechanisms, these probabilities depend on covariates and/or the outcome. The linear predictors for fully and partially missing outcomes were constructed as:
\begin{align}
\label{eq_alpha}
   \eta^{\text{Missing}}_i &= \alpha_0 + \alpha_1  \text{Age}_i + \alpha_2 \text{Sex}_i + \alpha_3 y_i, \\
\label{eq_beta}
\eta^{\text{Partial}}_i &= \beta_0 + \beta_1  \text{Age}_i + \beta_2 \text{Sex}_i + \beta_3 y_i.
\end{align}

We selected values for $\alpha_0, \alpha_1, \alpha_2$ and $\alpha_3$ as in Table \ref{MDM} to induce missingness according to the five mechanisms. Three additional MAR-Age mechanisms with varying quantities of missing data were also explored; details are in Supplementary File 2. \\
 
Probabilities for each missingness category were then obtained via the softmax function, i.e., a generalization of logistic regression to handle outcomes with three or more discrete categories:
\begin{align*}
\mathbb{P}(S_i= \mbox{Complete}) &= \frac{1}{1 + e^{\eta^{\text{Partial}}_i} + e^{\eta^{\text{Missing}}_i}}, \\
\mathbb{P}(S_i= \mbox{Partial})&= \frac{e^{\eta^{\text{Partial}}_i}}{1 + e^{\eta^{\text{Partial}}_i} + e^{\eta^{\text{Missing}}_i}}, \\
\mathbb{P}(S_i= \mbox{Missing}) &= \frac{e^{\eta^{\text{Missing}}_i}}{1 + e^{\eta^{\text{Partial}}_i} + e^{\eta^{\text{Missing}}_i}}.
\end{align*}

\begin{table*}[ht]
\centering
\renewcommand{\arraystretch}{1.5}
\begin{tabular}{|l|l|l|l|}
\hline
\textbf{Mechanism} & $\bm{\alpha}$ & $\bm{\beta}$ & \makecell[l]{\textbf{Missing Data Status} \\ \textbf{(\%Obs, \%Partial, \%Missing)}} \\ \hline

MCAR & $[-3.455, 0, 0, 0]$ & $[-3.86, 0, 0, 0]$ & Overall: (95\%, 2\%, 3\%) \\ \hline

\multirow{3}{*}{MAR Age} & \multirow{3}{*}{$[-1, -2, 0, 0]$} & \multirow{3}{*}{$[-1, -1.5, 0, 0]$} & Overall: (86\%, 8\%, 6\%) \\ \cline{4-4}
 & & & For Age=0: (59\%, 20\%, 20\%) \\ \cline{4-4}
 & & & For Age=1: (88\%, 7\%, 4\%) \\ \hline

\multirow{3}{*}{MAR Sex} & \multirow{3}{*}{$[-2.5, 0, 0.8, 0]$} & \multirow{3}{*}{$[-2.5, 0, 0.5, 0]$} & Overall: (83\%, 8\%, 9\%) \\ \cline{4-4}
 & & & For males: (86\%, 7\%, 7\%) \\ \cline{4-4}
 & & & For females: (77\%, 10\%, 13\%) \\ \hline

\multirow{5}{*}{MAR Age Sex} & \multirow{5}{*}{$[-0.5, -2, 0.5, 0]$} & \multirow{5}{*}{$[-1, -1, 0.2, 0]$} & Overall: (77\%, 13\%, 10\%) \\ \cline{4-4}
 & & & For Age=0: (45\%, 19\%, 36\%) \\ \cline{4-4}
 & & & For Age=1: (80\%, 12\%, 8\%) \\ \cline{4-4}
 & & & For males: (79\%, 12\%, 9\%) \\ \cline{4-4}
 & & & For females: (72\%, 13\%, 15\%) \\ \hline

\multirow{6}{*}{MNAR Age} & \multirow{6}{*}{$[-0.1, -1.5, 0, -0.018]$} & \multirow{6}{*}{$[-0.1, -0.5, 0, 0.01]$} & Overall: (75\%, 20\%, 5\%) \\ \cline{4-4}
 & & & For Age=0: (62\%, 25\%, 13\%) \\ \cline{4-4}
 & & & For Age=1: (76\%, 20\%, 4\%) \\ \cline{4-4}
 & & & For $y=0$: (54\%, 36\%, 10\%) \\ \cline{4-4}
 & & & For $0<y<70$: (70\%, 23\%, 6\%) \\ \cline{4-4}
 & & & For $y \geq 70$: (76\%, 19\%, 5\%) \\ \hline

\end{tabular}
\caption{Details of the five missing data mechanisms in the simulation study: values of parameters $\bm{\alpha}$ and $\bm{\beta}$ in Equations \eqref{eq_alpha} and \eqref{eq_beta}, and description.} 
\label{MDM}
\end{table*}

When the ordinal random variable $S_i$ takes value \textit{Missing}, the entire post hospital discharge period is missing. When $S_i$ takes value \textit{Partial}, we assume that a proportion of the $w_i$ days following hospital discharge is missing so that, over the period of interest, $w_i^{obs}$ days are observed and $w_i^{mis}$ are missing, with $w_i=w_i^{obs}+w_i^{mis}$. We generate the proportion of missing days per participant using a $Beta(2, 4)$ distribution, which has expectation of 0.33. We assume that, amongst $w_i^{obs}$ observed post-discharge days  $y_{R_i}^{obs} \in \left\{ 0, ...,w_i^{obs} \right\}$ days in readmissions are observed among $y_R$ total days in reaadmission, where 
\begin{equation}
    y_{R_i}^{obs} \sim \mbox{Hypergeometric}(w_i, y_{R_i} ,w_i^{obs}). 
\end{equation}
Under the MAR missingness mechanisms, this assumes that days at home and readmissions are equally likely to be missing, which may be unrealistic in practice. In contrast, for the MNAR-Age assumption, the $w_i^{mis}$ were deterministically selected to have as many hospital days as possible. \\

We compared the following methods for handling missing data. For each MI method, imputation proceeded separately by trial arm with $M=35$ imputations. 

\begin{itemize}
    \item \textbf{Complete case analysis}, where analysis is restricted to participants who have all components fully observed; 
    \item \textbf{Complete case analysis (derived)}, where DAH90 is derived for participants who die but have missing location data in the post hospital discharge period. 
    \item \textbf{MI-composite}: If there is at least one day where location is missing, the entire DAH90 observation $y_i$ is set to missing, irrespective of the death status, and the response $y_i$ is imputed via MI with key covariates (which affect the missingness mechanism).  
    \item \textbf{MI-components}: If there is at least one day where location is missing in the post hospital discharge period, $y_{R_i}$ is set to missing and imputed via MI with key covariates and $y_{E_i}$, the extended initial hospital stay for participant $i$. The imputed DAH90 outcome is then obtained via Equation \eqref{DAH90_formula}.
    \item \textbf{MI-components (binomial)}: This approach makes use of partially observed post discharge data. If $y_{R_i}^{obs}$ readmissions are observed in a period of $u_1^{obs}$ days and there are $u_1^{mis}$ days missing, we assume that the number of readmissions within $u_1^{mis}$ days can be generated from a binomial distribution with success probability depending on key covariates and $y_{E_i}$. The total number of readmissions is then given by  $y_{R_i}^{imp}+y_{R_i}^{obs}$, and the imputed DAH90 outcome is obtained via Equation \eqref{DAH90_formula}.
\end{itemize}

For MI-composite and MI-components, we considered both parametric regression (``norm" in \texttt{mice}) and predictive mean matching (``pmm" in \texttt{mice}) with 5 or 10 as the number of donors. In cases where deaths have missing location values, components were included in the imputation procedures, and the final DAH90 value was assigned the value of zero.\\

The primary analysis of the trial was a Mann-Whitney-Wilcoxon test. The estimand is the probabilistic index: 
\begin{equation}
\label{theta_estimand}
    \theta = \mathbb{P} \left(Y^1 > Y^0 \right) + \frac{1}{2} \mathbb{P} \left(Y^1=Y^0 \right),
\end{equation}

where $Y^{1}$ and $Y^{0}$ respectively denote a random draw from the distribution of outcomes for arm 1 and 0. We also computed the difference in medians between the two treatment groups. \\

The Mann-Whitney-Wilcoxon test is performed on each of the $M=35$ imputed datasets, and the median p-value is obtained. This analysis approach is taken to mirror the NOTACS Statistical Analysis Plan (SAP), ensuring that the simulation study remains directly applicable to the trial's intended analysis.\\

We repeated the simulation 10,000 times and summarised the output using the following performance measures: 

\begin{enumerate}
    \item Type I error rate (Under the Null scenario);
    \item Power (Under the Alternative scenario);
    \item Mean of $\theta$;
    \item Mean of difference in medians between the two groups.
\end{enumerate}

Simulations were performed in \texttt{R} version 4.5.1 using the \texttt{R} package \texttt{mice}.

\section{Results}

\begin{figure*}[] 
    \centering
    \includegraphics[width=0.8\textwidth]{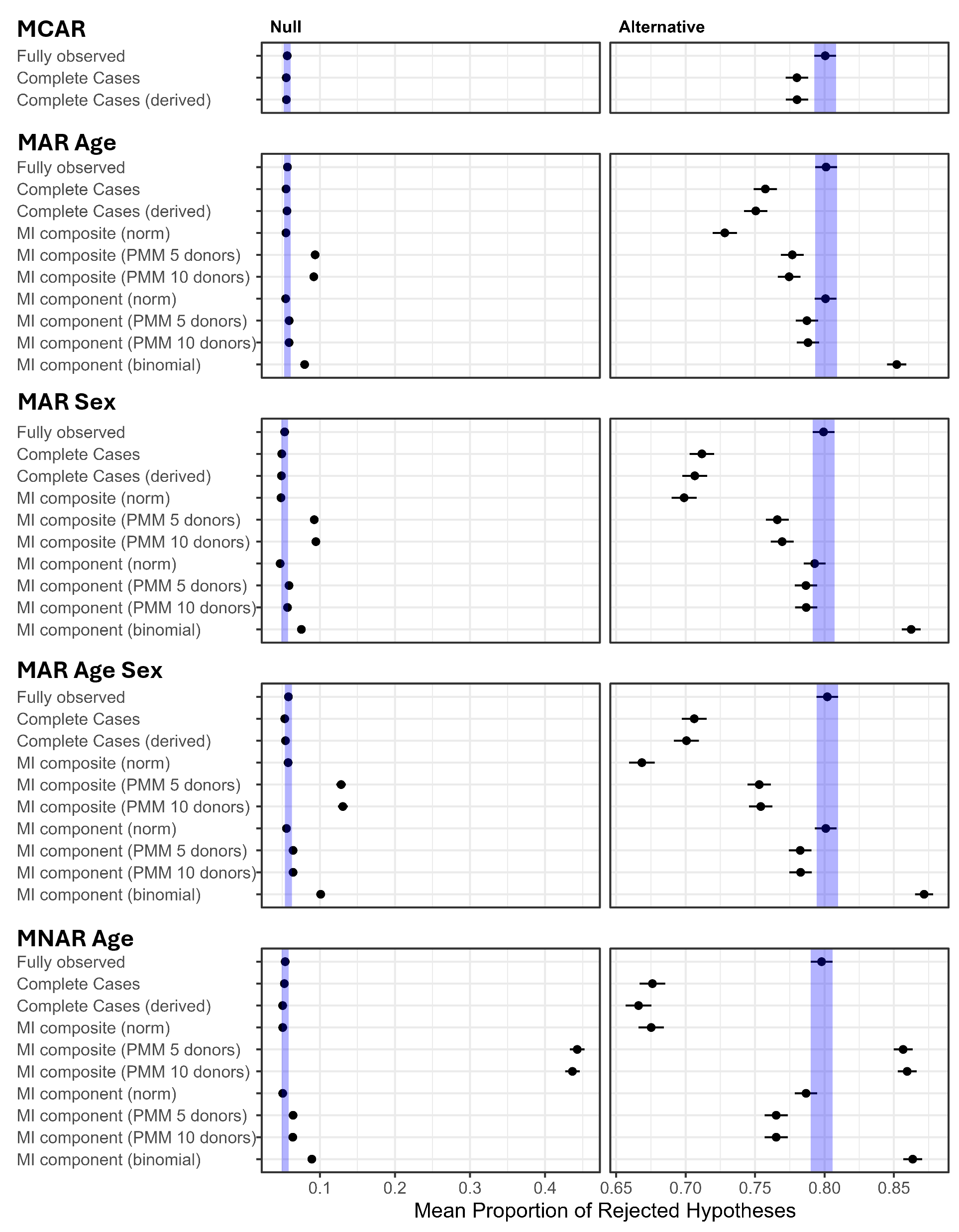}
    \caption{Simulation results: Type I error (left) under the Null scenario and Power (right) under the Alternative scenario are shown. The performance measure is indicated with $\pm 1.96 \times $ Monte Carlo error bars. Results when data are fully observed are displayed with extended bands;  confidence intervals which do not overlaps with these extended bands show difference in performance that are beyond MC error.}
    \label{Rejected_Null}
\end{figure*}

Figure \ref{Rejected_Null} displays results in terms of type I error under the null (left) and power under the alternative (right) across the five missingness mechanisms described in Table \ref{MDM} and for several missing data strategies. When data are MCAR (5\% missing), using complete case analysis does not affect the type I error but leads to slightly reduced power, whether or not DAH is derived for patients who die. For MAR-Age ($14\%$ missing), all methods lead to a type I error close to the 5\% nominal value except for the `MI composite (PMM)' and `MI component (binomial)' approaches. Among the methods leading to the target type I error, all methods except `MI component (norm)' lead to some decrease in power, where this decrease is smaller for the `MI component' approaches. Results were similar across the other MAR settings considered here. \\

Figure 1 in Supplementary File 2 displays results for the MAR-Age scenario with different quantities of missingness. For example, when $25\%$ of outcomes are missing, the type I error rate increases to $32.2\%$, illustrating a very large consequence as the proportion of missing data increases. Figures 2 and 3 in Supplementary File 2 provide additional results to illustrate why type I error inflation occurs; namely, the imputation model on the composite level is misspecified and can lead to imputation of value in the clump-near-zero with high values. Under the null, this may occur more in one arm than another, driving the type I error inflation. \\

In the MNAR-Age scenario, we observe that both `MI composite (PMM)' and `MI component (binomial)' appear to be problematic with type I inflation and to an increase in power driven by the false positives induces by the lack of type I error control. \\

Supplementary File 2 displays results for $\hat{\theta}$ as well as the estimated difference in medians between the two treatment groups. Briefly, we note that:
\begin{itemize}
    \item Complete Cases (derived) appears not to lead to biases in this setting, likely due to the low proportion of deaths. In most settings, deriving deaths is not recommended \citep{Pham2021};
    \item While MI-components (norm) has good performance in terms of testing, it appears to lead to greater biases, both for $\hat{\theta}$ and for the difference in medians, compared to MI-components (PMM).
\end{itemize}

\section{Discussion}
DAH is a composite outcome that is gaining traction in postoperative trial designs \citep{YuanPLACEHOLDER}, but its complex distributional characteristics make handling missing data challenging. Naive approaches such as imputing on the composite level with PMM, is not only inefficient -- as it discards data from the fully observed initial hospital stay -- but also leads to type I error inflation. Our simulation study motivated by the specific setting of the NOTACS trial shows that MI-composite (PMM) can lead to type I error inflation that can reach $32\%$ when $25\%$ of outcomes are missing. This is likely driven by the misspecification of the composite-level imputation model. Our simulations also demonstrate that MI on the component level, either parametrically or via PMM, leads to good properties in terms of type I error and power.  \\

There are several key areas of future work. First, our investigation focused on the Mann-Whitney-Wilcoxon test and the median p-value was reported following MI. The median p-value approach has been shown to lead to type I error inflation in an extensive simulation study by \cite{Austin2025}. While our simulation study found that type I error inflation occurs specifically when the composite outcome is imputed via PMM, the median p-value method may also contribute to type I error inflation if larger proportions of missing data were observed. Combining results from MI, for example via Rubin's Rules for the probability index, is an area of future work. Second, Exploration of missing data methods for other analysis methods for DAH are warranted, such as quantile regression \citep{Chung2025, Myles2018}. Third, imputation on the component level with ``norm'' or PMM discards the partial data in post hospital discharge period. While MI-components (binomial) retains the partial data, we found that it did not handle the skewed distribution of readmissions well. Therefore, future work could explore improved approaches to imputing the missing part of the post hospital discharge period while retaining the observed part. Fourth, as readmissions are less likely to be missing than days at home, sensitivity analyses to the MAR assumption are needed. \\

Finally, this investigation used a simplified binary classification for DAH, distinguishing only between ``home'' and ``hospital''. The definition used in the NOTACS trial was more granular and excluded any days spent in an escalated state of care (e.g. nursing home for a participant who ordinarily resided at home). 
Future research should explore imputation approaches consistent with the granular definition of DAH.\\


\section*{Declaration of Conflicts of Interest}
SSV is on the advisory board for PhaseV (unrelated to this work).

All other authors have no conflicts of interest to declare.

\section*{Funding}
MST, Advanced Fellow, NIHR305417, is funded by the National Institute of Health and Care Research for this research project. The views expressed are those of the authors and not necessarily those of the NIHR or the Department of Health and Social Care. 

LY is supported by the Medical Research Council Trial Methodology Research Partnership (TMRP) Doctoral Training Partnership (DTP) (grant number MR/W006049/1).

SD is funded by the Papworth Trials Unit Collaboration.

\section*{Supplementary Materials}
Supplementary File 1 provides further detail on the model for DAH90 by \cite{YuanPLACEHOLDER}. Supplementary File 2 provides additional simulation results. \texttt{R} code to reproduce simulations are provided in the GitHub repository for this project: \href{https://github.com/mst1g15/Missing_DAH/}{https://github.com/mst1g15/Missing\_DAH/}. 

\newpage

\bibliography{library}

\end{document}